\documentclass[usenatbib,useAMS,twocolumn]{mn2e}
\usepackage{psfig}
\usepackage{ltxtable}
\usepackage[latin1]{inputenc}
\newcommand{\dpart}[2]{\frac{\partial #1}{\partial #2}}
\newcommand{\dth}[3]{\dpart{#1}{#2}{\big \rfloor}_{#3}}

\def\pmb#1{\setbox0=\hbox{#1}
  \kern-.025em\copy0\kern-\wd0
  \kern.05em\copy0\kern-\wd0
  \kern-.025em\raise.0433em\box0 }

\title[C-flash and SN Ia ignition]
{The C-flash and the ignition conditions of type Ia supernovae}

   \author[P. Lesaffre, Z. Han, C. A. Tout, P. Podsiadlowski \& R. G. Martin]
{P. Lesaffre,$^{1,2}$\thanks{lesaffre@ast.cam.ac.uk}      
	Z. Han, $^{3}$
	C. A. Tout, $^{1,4}$
        Ph. Podsiadlowski$^{2}$
	and R. G. Martin$^{1}$\\
$^{1}$Institute of Astronomy, Madingley Road, Cambridge 
CB3~0HA, UK \\
$^{2}$University of Oxford, Department of Astrophysics,
 Oxford OX1~3RH, UK\\
$^{3}$National Astronomical Observatories/Yunnan Observatory, 
the Chinese Academy of Science, PO Box 110, Kunming 650011, China\\
$^4$Centre for Stellar and Planetary Astrophysics, School of
Mathematics, Monash University, Clayton, Victoria 3800, Australia
}
\begin{document}

   \date{Received September 15, 1996; Accepted March 16, 1997}

   \maketitle

\begin{abstract} 
  Thanks to a stellar evolution code able to compute through the
  C-flash we link the binary population synthesis of single degenerate
  progenitors of type Ia supernovae (SNe~Ia) to their physical
  condition at the time of ignition. We show that there is a large
  range of possible ignition densities and we detail how their
  probability distribution depends on the accretion properties. The
  low density peak of this distribution qualitatively reminds of the
  clustering of the luminosities of Branch-normal SNe~Ia. We tighten
  the possible range of initial physical conditions for explosion
  models: they form a one-parameter family, independent of the
  metallicity.  We discuss how these results may be modified if we
  were to relax our hypothesis of a permanent Hachisu wind or if we
  were to include electron captures.
\end{abstract}

\section{Introduction}

  Phillips relations \citep{P93} have made type Ia supernovae (SNe~Ia)
  a very useful tool for cosmology. The correlation between their
  absolute peak brightness ($M_{\rm peak}$) and their decline rate
  ($\Delta m_{15}$) has been used to infer distances of high redshift
  galaxies and refine our knowledge of cosmological parameters
  \citep[e.g.][]{T01}. However current models are not able to
  reproduce the range of $\Delta m_{15}$ observed: we still do not
  understand what primary parameter is responsible for the main
  diversity of SNe~Ia.  Moreover there exists a finite spread of
  $M_{\rm peak}$ for a given $\Delta m_{15}$ which hints at the
  existence of a secondary parameter.  Knowledge of both these
  controlling parameters could help remove any bias from cosmological
  measurements due to an evolution of the secondary parameter and help
  reduce the scatter of the Phillips relation.

  With synthetic light curves and spectra with explosion models,
  attempts have been made to infer what parameters of the explosion
  control which observational property \citep[][ for
  example]{RH04}. In this paper we proceed from the binary population
  synthesis of SNe~Ia progenitors where we only consider the single
  degenerate channel, one of the most promising channels for the
  progenitors of SNe~Ia. We model their evolution until we get the
  physical conditions at the time of the ignition. We start our
  computations just after a white dwarf (WD) has been formed through a
  common envelope evolution phase. The initial parameters are the WD
  composition, its initial mass $M_{\rm WD}^{\rm i}$, the initial mass
  $M_2^{\rm i}$ of the secondary and the initial separation $a_{\rm
  i}$.  The WD cools down while the secondary evolves until it
  overflows its Roche lobe at the end of its main-sequence evolution
  (cooling phase). Accretion on to the WD heats it up and enhances its
  central density until the carbon fusion starts (beginning of the
  accretion phase). The rate of energy production due to burning and
  accretion soon overcomes the losses due to escaping neutrinos and
  photons: thermal balance does not hold anymore (birth of the
  convective core). However, the high electron degeneracy and the high
  temperature sensitivity of the burning rate prevent expansion from
  controlling the burning and the convective core has to grow very
  fast to cope with the increasing energy generation rate: a flash
  ensues (C-flash phase).  Convective flows are able to get rid of the
  energy released by carbon burning until the turnover time scales
  become too long relative to the heating rate. At this point it is
  believed that one or multiple bubbles ignite and a flame front
  propagates out and eventually unbinds the star (ignition or
  beginning of the explosion).  In the present work we investigate the
  probability distributions and correlations between various
  characteristics of the star at the time of ignition which are
  thought to influence the outcome of the explosion.

  Section~2 presents the details of our numerical setup and the main grid of
  models. In section~3 we describe the resulting ignition conditions.
  In sections 4 and~5 we investigate the role of metallicity and the
  effects of binary evolution respectively.  And in sections 6 and~7
  we discuss and summarise our results.

\section{Method}

\subsection{Stellar evolution code}

  We improved the Eggleton code \citep{E71,P95} to enable the
  modelling of white dwarfs from their cooling tracks and the start of
  the accretion to the very late phases of the C-flash until shortly
  before the explosion takes place. Note that neutrino losses in the
  Eggleton code are from \citet{I92}: they are an important factor
  that determine the position of the central density-temperature
  tracks of the WD.

  We consider only the $^{12}$C+$^{12}$C fusion reaction and assume
  that $^{24}$Mg is its only product. We hence completely neglect all
  nucleosynthesis issues such as Urca processes which are likely to
  occur during the C-flash \citep[see][]{L05a}.

  For each chemical species described in the code we solve not only
  for its mass fraction but also for its spatial gradient. This allows
  us to compute the mixing for realistic diffusion time scales as
  given by the mixing length theory (MLT). This is important because
  close to ignition convective turnover time scales become comparable
  and even longer than the burning time scales.

  We compute the equations of hydrostatic stellar evolution on a
  staggered mesh which allows us to take arbitrarily small time steps
  required by the burning time scales. This is also needed to
  stabilise chemical equations including both gradients and
  composition. Radius, mass, luminosity and chemical gradients are
  hence computed on the interfaces of the computational shells whereas
  all other variables are taken at the centre of mass of each zone.

  We make use of a moving grid algorithm \citep{DD87,L04} with a time
  delay of one year. This friction of the grid limits numerical
  diffusion and appears to stabilise the code against rapid changes at
  the onset of accretion and near ignition.

  Last but not least we implemented a scheme to track the boundaries
  of convective regions within a zone. As the time scale for burning
  decreases, a discrepancy builds up between the rate of change of
  temperature within and outside the convective core. When convective
  boundaries are tied up to the interfaces of the computational zones,
  the code has more and more difficulty to converge and breaks down
  way before the explosion takes place. We need to release the
  convective boundaries and interpolate the convective
  criterion\footnote{We use the Schwarzschild criterion without
  overshooting} to infer their position inside a zone. We are then
  able to account properly for the discontinuity in the rate of change
  of temperature. Chemical rates of change benefit from exactly the
  same treatment but chemical profiles are discontinuous at convective
  frontiers (unlike the temperature) and they require an additional
  flux through the advancing boundary of the core.
  
\subsection{Initial WD models}

  We build a series of WD models of different masses $M^{\rm i}_{\rm
  WD}$ by accreting on a very low-mass (0.3~M$_\odot$) $\rm C+O$~WD
  model at a rate of 10$^{-7}\,$M$_\odot$/yr which mimics the growth
  of the CO core during the giant branch evolution. This gives a very
  crude estimate for the thermal state of the WD when it is first
  formed. This is however not essential as the thermal profile relaxes
  during the subsequent cooling phase. \cite{U99} and \cite{D01} have
  shown how the mass and metallicity of the progenitor star of the WD
  influence its chemical state.  In the present study we use their
  results to build synthetic chemical profiles of the initial white
  dwarf. We use their central values for the C/O ratio (linearly
  interpolated at $M^{\rm i}_{\rm WD}$) to build a central core of
  mass half the white dwarf. This is then connected linearly to a C/O
  ratio of 1 uniform in the outermost 10\% in mass of the WD.

\subsection{Cooling phase}

\begin{figure}
\centerline{
\psfig{file=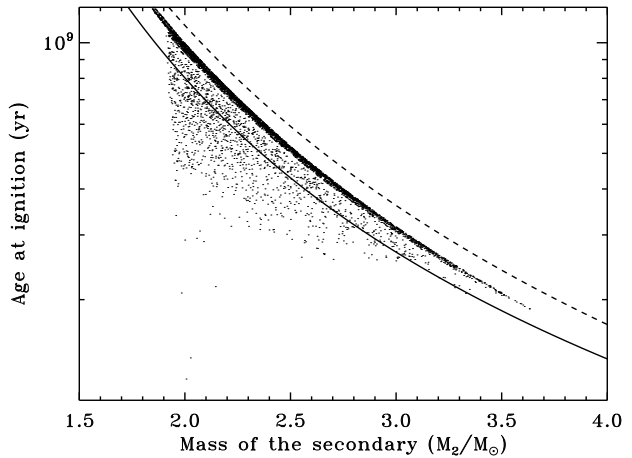,angle=90,width=9cm}
} \caption{Age vs. mass of the secondary for our sample of type Ia
progenitors. The solid and dashed lines are analytical estimates for the age of
the main sequence taken from \citet{E89} (solid) and \citet{H00} (dashed).}
\label{agem2}
\end{figure}

  The newly formed WD cools down at constant mass until the secondary
  star fills its Roche lobe, which generally happens at the end of its
  main sequence, on its way to become a giant (in the Hertzsprung gap)
  or on the giant branch. In each case the cooling time is close
  to the main-sequence age of the secondary star (see figure
  \ref{agem2}). For a given WD mass the length of the cooling phase
  generally determines the temperature at its centre.  It does not
  depend much on its initial thermal profile.  Hence the initial conditions
  for the accretion phase depend only on the age of the
  secondary and the mass of the WD.

\subsection{Accretion phase}

  When the secondary starts to overflow its Roche lobe it transfers mass
  on to the WD. How fast an accretion rate the WD can cope with is a
  crucial question and is still the object of much debate. In this
  study we follow \cite{HP04} and assume that the Hachisu wind model
  \citep{H96} holds. The WD accommodates high mass transfer rates by
  getting rid of the extra mass through a wind as long as the mass
  transfer rate exceeds a critical rate $\dot{M}_{\rm cr}$ given by
\begin{equation}
\label{mcr}
\dot{M}_{\rm cr}=5.3 \times 10^{-7}\frac{(1.7-X)}{X}
(M_{\rm WD}/{\rm M}_\odot -0.4)\,{\rm M}_\odot\,{\rm yr}^{-1}
\end{equation}  
  where $X=0.7$ is the hydrogen mass fraction in the accreted material
  from the secondary and $M_{\rm WD}$ is the mass of the WD.  As a
  further simplification we assume that the mass transfer rate is {\it
  always} higher than this critical rate so that the Hachisu wind
  determines the accretion rate on to the WD.  The net growth of the
  C+O WD is finally modulated by the efficiency of He-shell flashes
  according to
\begin{equation}
\dot{M}_{\rm WD}=\eta_{\rm He}(\dot{M}_{\rm cr})\times\dot{M}_{\rm cr}
\end{equation}
  where $\eta_{\rm He}$ is given by equation (4) of \cite{HP04}.  With
  these assumptions the growth rate of the WD depends only on the mass
  of the WD. We estimate the effects of the end of the wind phase in
  section~5.

\subsection{Ignition}

  When convection is no longer able to evacuate the energy input from
  the C-burning, thermal runaway starts in some bubbles near the
  centre and the star explodes. Exactly how and when this occurs is
  still unclear but recent developments allow us to pinpoint analytic
  conditions for the ignition of the flame \citep{W04,WW04}. Here, we
  stop our computation when the differential burning time scale
  $t_{\rm b}$ is shorter by a given fraction $\alpha$ of the local
  convective turnover time scale $t_{\rm c}$ at some point in the
  star.  Here $t_{\rm b}$ is the e-folding time scale of a temperature
  fluctuation subject to C-burning:
\begin{equation}
t_{\rm b}^{-1}=\frac{1}{c_{\rm P} T}\dth{q}{\ln T}{\rho},
\end{equation}
  where $T$ is the temperature, $\rho$ is the mass density, $c_{\rm
  P}$ is the specific heat at constant pressure and $q$ is the rate of
  energy generation due to C-burning and $t_{\rm c}$ is a convective element's
  crossing time over a pressure scale height:
\begin{equation}
t_{\rm c}=\frac{H_{\rm P}}{u_{\rm c}},
\end{equation}
  where $H_{\rm P}$ is the pressure scale height and $u_{\rm c}$ is the
  convective velocity as given by MLT.

  When $t_{\rm b}=t_{\rm c}$ ($\alpha=1$) the differential burning
  time scale is of the same order as the convective turnover time
  scale.  At this point differential reactivity (the fact that upward
  and downward moving fluid elements burn at different rates) should
  be included in the treatment of convection to get accurate mixing
  and temperature excesses \citep[see][]{L05a,L05b}.

  In our models (in which we use classical MLT) it turns out that the
  temperature excess is typically 0.025\% when this happens. It takes
  roughly eight e-foldings to grow from 0.025\% to 100\%.  Hence can
  we estimate that temperature fluctuations will be big for $t_{\rm
  b}=t_{\rm c}/8$ ($\alpha=1/8$). At this point, a typical ascending
  temperature fluctuation increases its temperature up to twice the
  average temperature before it is mixed.  We take this as a plausible
  point for the start of the thermal runaway.

  \citet{WW04} consider a temperature background fixed in time and
  analyse the growth of perturbations relative to this background. As
  a result they find that the relevant time scale for the growth of
  the temperature fluctuations is the e-folding time scale for the
  overall burning rate
\begin{equation}
t_{\rm B}^{-1}=\frac{q}{c_{\rm P} T} \mbox{.}
\end{equation}
  Because of the high power exponent in the temperature dependence of
  the burning rate $t_{\rm b}\simeq t_{\rm B}/22$.  Hence in this picture
  $\alpha=1/22$ is a more relevant requirement for the start of
  the explosion.  Note that $t_{\rm B}$ also indicates the time scale
  of variation for the average temperature in the centre of the
  star. When $t_{\rm B}<t_{\rm c}$ a time-dependent convective model
  should be used because the turnover time scales become longer than
  the global evolutionary time scales.

  To account for the uncertainty in what determines the start of the
  explosion, we record models when $\alpha$=1,1/8 and 1/22 as possible
  times for the start of the explosion. In practice all our
  simulations reach the $\alpha=1$ point, half of them reach
  $\alpha=1/8$ but all break down before $\alpha=1/22$. We hence
  extrapolate our results from our last converged model to get the
  conditions for all these different possibilities for the ignition
  point.  In almost all our simulations the criterion for explosion is
  first fulfilled at the centre because the local convective time
  scale is longer there. The only exception is for one run which
  started off-centre carbon burning and the convective zone never
  reached the centre (its parameters are $t_{\rm a}=0.8\,$Gyr and
  $M^{\rm i}_{\rm WD}= 0.9\,$M$_\odot$, the bottom of the convective
  region is at $0.05\,$M$_\odot$ at ignition). We display results for
  this run only in figure \ref{rhoc} as most of the quantities that
  take part in the correlations are defined at or relative to the centre.

\subsection{Parameter space}

  Because we always assume a Hachisu wind phase (except in section~5),
  the entire evolution of the WD is determined by the cooling age and
  the initial mass of the WD. Starting with a given initial model, we
  let it cool down for a time $t_{\rm a}$ after which we trigger the
  accretion rate $\dot{M}_{\rm WD}$. The beginning of the accretion is
  smoothed linearly with a short time scale (10$^5\,$yr) to mimic the
  rise observed in \cite{HP04}.  We build a grid of 24 models spanning
  six different masses $M_{\rm WD}^{\rm i}$ evenly distributed in the
  range [0.7,1.2]$\times$ M$_\odot$ and $t_{\rm a}$=0.1, 0.2, 0.4 and
  0.8~Gyr. This grid of models is designed to sample the
  parameter space of SNe~Ia progenitors found by \cite{HP04}. The
  metallicity is $Z=0.02$ and we use initial C/O ratios from
  \cite{U99} for this metallicity.

\section{Ignition conditions}

  Parameters which have been found to influence the explosion outcome
  are density, temperature, carbon mass fraction and convective
  properties.  We investigate here how they are related to the
  parameters $M^{\rm i}_{\rm WD}$ and $t_{\rm a}$.


\subsection{Central density}

\begin{figure}
\centerline{
\psfig{file=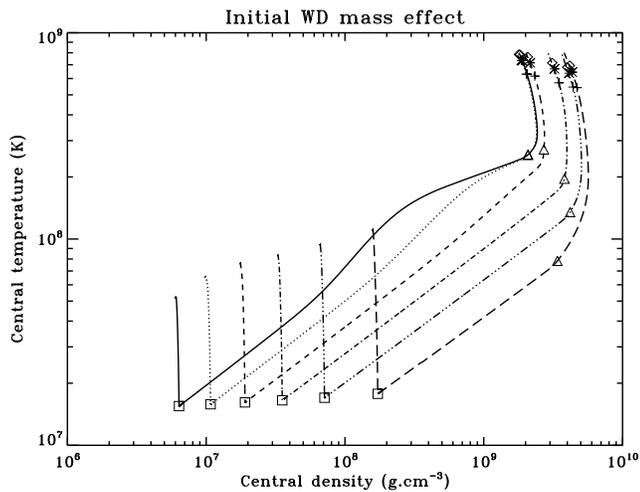,angle=90,width=9cm}
} \caption{Central density-temperature tracks for different initial WD
masses (from left to right: 0.7, 0.8, 0.9, 1.0, 1.1 and 1.2~M$_\odot$)
and the same cooling age $t_{\rm a}=0.4$~Gyr. Squares, triangles,
crosses, stars and diamonds indicate respectively the start of the
accretion phase, the growth of the convective core and the $\alpha=1$,
1/8 and 1/22 ignition points.  }
\label{rTmass}
\end{figure}

\begin{figure}
\centerline{
\psfig{file=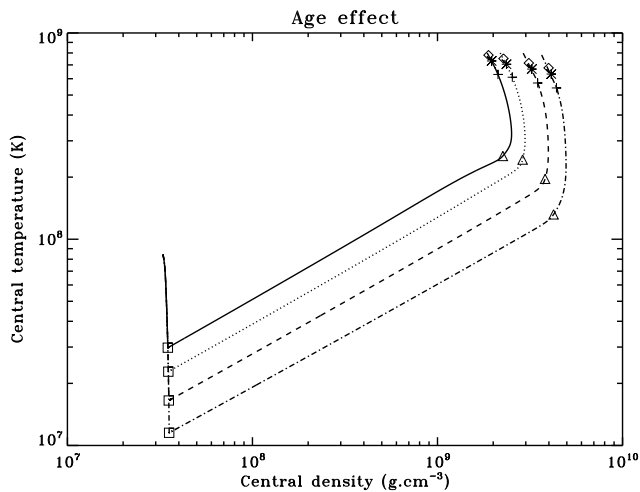,angle=90,width=9cm}
} \caption{Central density-temperature tracks for different cooling
ages (from left to right: 0.1, 0.2, 0.4 and 0.8~Gyr) and the same
initial WD mass 1~M$_\odot$. Symbols are as in figure~\ref{rTmass} }
\label{rTage}
\end{figure}

  Larger initial masses $M^{\rm i}_{\rm WD}$ yield higher ignition
  densities because their initial densities are higher (see figure
  \ref{rTmass}).  A longer $t_{\rm a}$ yields a lower initial
  temperature and a hence higher ignition density (see
  figure~\ref{rTage}).

\begin{figure}
\centerline{
\psfig{file=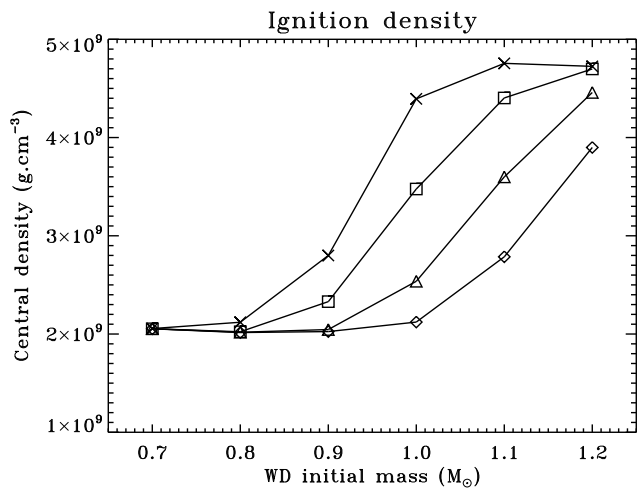,angle=90,width=9cm}
} \caption{Ignition densities (defined by $\alpha=1$) vs. initial WD
mass for different cooling ages $t_{\rm a}=10^8\,$yr (diamonds),
$t_{\rm a}=0.2$~Gyr (triangles), $t_{\rm a}=0.4$~Gyr
(squares) and $t_{\rm a}=0.8$~Gyr (crosses).
Densities are only slightly lower for other values of $\alpha$.}
\label{rhoc}
\end{figure}

  The dependence of density on $M_{\rm WD}$ and $t_{\rm a}$ is
  summarized in figure~\ref{rhoc}.  A WD with a smaller $M^{\rm
  i}_{\rm WD}$ has a shorter thermal diffusion time scale and a longer
  accretion phase.  As a result heat waves have time to propagate
  through the star and the thermal profile of the star reflects the
  {\it global} balance between the energy input from the accretion
  (and burning) and the energy losses from neutrinos. In the rest of
  the paper we refer to this state as global thermal balance.  WDs in
  a state of global balance all end up on the same $\rho_{\rm
  c}$-$T_{\rm c}$ track because the accretion rate depends only on
  $M_{\rm WD}$. As a result, they all have the same ignition density.
  This is reflected in the distribution of ignition densities which
  has a spike at the lowest density.
 
  Densities also seem to all converge towards the same maximum density
  for higher initial masses $M^{\rm i}_{\rm WD}$. This is due to the
  shape of the ignition curve which is almost vertical in the $\rho-T$
  plane because electron screening enhances C-burning at high
  density. As a result, the $\rho_{\rm c}-T_{\rm c}$ tracks that go to
  high densities tend to all start the C-flash phase at the same
  density.  Note also that these high densities point towards the
  importance of electron captures, not yet included in this study.

  Ignition density is the parameter that varies most along our grid of
  parameters. We shall show that most of the other physical properties
  at ignition are well correlated with this parameter.

\subsection{Ignition Temperature}

  A lower $\alpha$ yields hotter and only slightly less dense ignition
  conditions because matter is still degenerate at this point.
  We now examine how the temperature varies according to density when
  we define the ignition conditions following a given $\alpha$.

\begin{figure}
\centerline{
\psfig{file=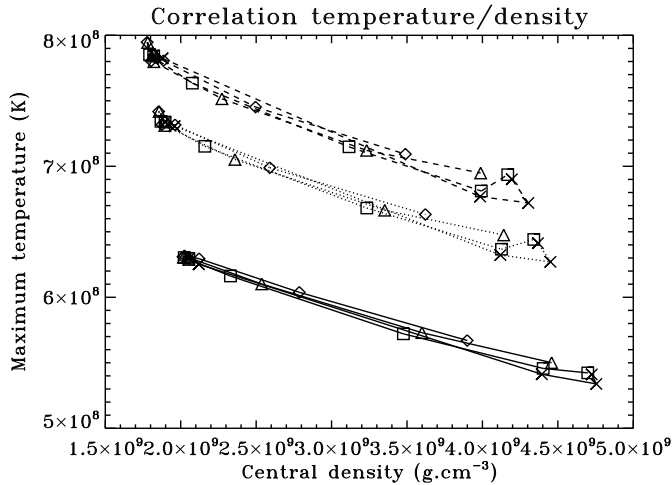,angle=90,width=9cm}
} \caption{Central temperature vs. central density at ignition central
density for $\alpha=1$ (solid), 1/8 (dotted) and 1/22 (dashed). Curves and
signs are otherwise labelled as in figure~\ref{rhoc}.}
\label{Tc}
\end{figure}

  The convective turnover time scale $t_{\rm c}$ is rather independent of
  any other parameter at the time of ignition as shown in section
  \ref{convection}. By contrast the electron screening corrections to
  the rate of C burning increase it at high density. As a result,
  because the ignition temperature is defined by the relation
  $t_{\rm b}=\alpha t_{\rm c}$, ignition temperatures are lower for higher
  densities (see figure~\ref{Tc}). This might introduce interesting
  effects in light of the results of \cite{I05} who find a strong
  sensitivity to the temperature for the properties of igniting
  bubbles.

\subsection{Convection}
\label{convection}

\begin{figure}
\centerline{
\psfig{file=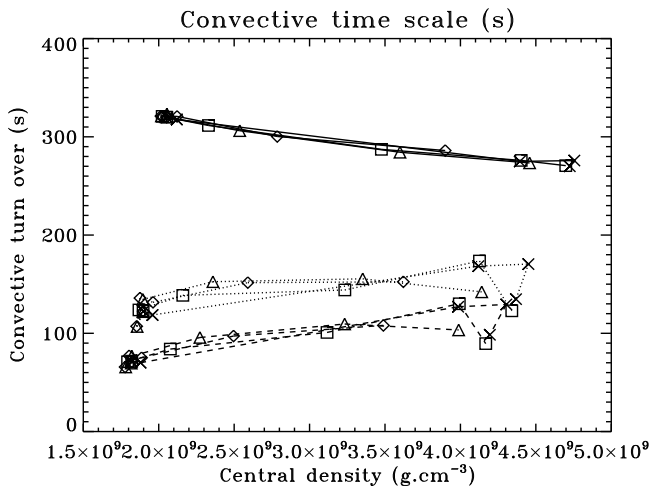,angle=90,width=9cm}
} \caption{Central convective time scale $t_{\rm c}$ vs. central density at
ignition, labels as in figure~\ref{Tc}.  }
\label{tconv}
\end{figure}

  Explosion models have recently focused on the importance of the
  convective state at the beginning of the explosion
  \citep{R02,HS02,GB05,WW04}.  It is interesting to note that our
  simulations (based on MLT) show very little variation for the
  convective state at the time of ignition.  The turnover time scale
  $t_{\rm c}$ appears to be more or less independent of $t_{\rm a}$ and $M_{\rm
  WD}^{\rm i}$. Naturally these time scales depend upon our definition for
  the ignition as $t_{\rm c}$ directly enters our criterion (see figure
  \ref{tconv}). We recover values in the range 10-100~s as quoted by
  \cite{WW04} when we use $\alpha=1/22$.  Also worth noting is a weak
  dependence of $t_{\rm c}$ on central density. This could be enhanced by
  convective Urca processes due to the feedback of electron captures
  on convective velocities \citep[see][]{L05a}.  

\begin{figure}
\centerline{
\psfig{file=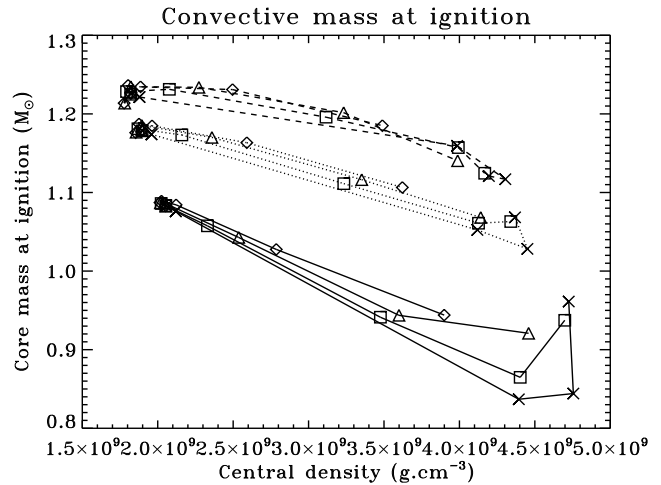,angle=90,width=9cm}
} \caption{Convective core mass at ignition vs. central density at
ignition, labels as in figure~\ref{Tc}.}
\label{mf}
\end{figure}

  The convective core mass at ignition is mainly determined by the
  central density which sets the pressure--mass profile and the
  central temperature which determines the extent of the convective
  core in the temperature--pressure profile. It is hence not
  surprising that we recover the same kind of correlations as for the
  central temperature (see figure~\ref{mf}). We investigate this
  parameter in light of the rotating progenitor models computed by
  \cite{YL04} who found super-Chandrasekhar solutions with a strongly
  differentially rotating layer around the mass shells
  $m=1.1\,$M$_\odot$ to $m=1.25\,$M$_\odot$.  If the convective core
  is able to reach this shell before ignition, it might suppress the
  centrifugal support and leave a higher than Chandrasekhar mass
  object which may then collapse instead of exploding. On the other
  hand convective instability might be suppressed by differential
  rotation and that would trigger the thermonuclear runaway sooner.
  Figure~\ref{mf} suggests that the convective core may sometimes
  reach the differentially rotating layer, depending on what
  definition we use for the ignition point.

\subsection{Carbon mass fraction}

\begin{figure}
\centerline{
\psfig{file=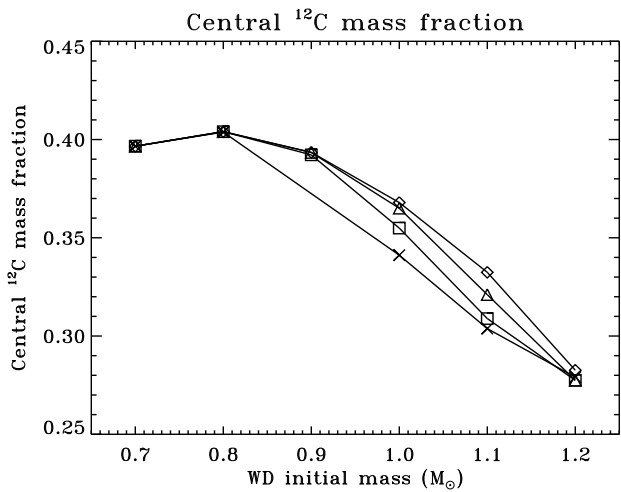,angle=90,width=9cm}
} \caption{Carbon central mass fraction at ignition vs. initial WD
mass, labels as in figure~\ref{rhoc}.  }
\label{xc}
\end{figure}

  The evolution of the central carbon mass fraction is mainly driven
  by the growth of the convective core which mixes material richer in
  carbon at a higher rate than the carbon is consumed at the
  centre. As a result the final carbon mass fraction depends
  essentially on the composition profile and the initial mass of the
  WD, that is on the initial $M^{\rm i}_{\rm WD}$ (see
  figure~\ref{xc}). The fact that there is little variation among
  different WDs and the recent result that the explosion might be less
  sensitive than we think to this parameter \citep{RH04} lead us to
  conclude that this might not be an essential parameter for
  understanding the observational properties of SNe~Ia. However it is
  the only parameter that slightly decouples from the central density.

\subsection{Magnesium mass fraction}

\begin{figure}
\centerline{
\psfig{file=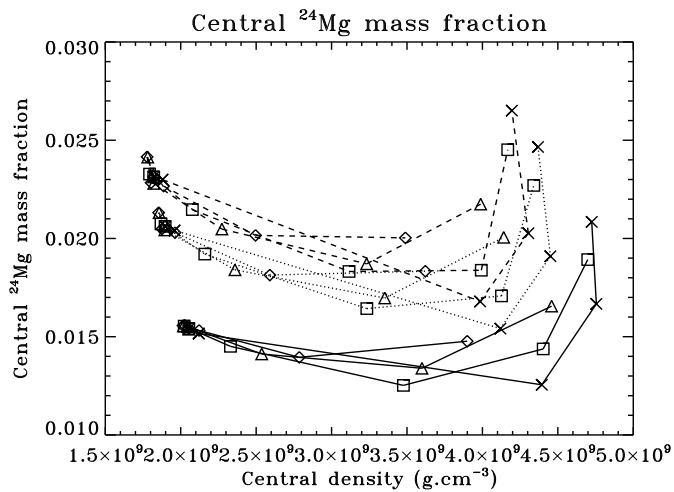,angle=90,width=9cm}
} \caption{Correlation between Magnesium mass fraction and central
density at ignition, labels as in figure~\ref{Tc}.}
\label{mgr}
\end{figure}

  Thanks to our oversimplified treatment of C-burning, the $^{24}$Mg
  mass fraction represents the amount of carbon burnt during the
  C-flash.  We show that this is weakly dependent on the central
  density even though the trend is not clear at high densities (see
  figure~\ref{mgr}). Later ignition (lower $\alpha$) naturally burns
  more carbon.  The amount of carbon burnt is of interest with regard
  to Urca species which are among the products of C-burning ($^{23}$Na
  and $^{25}$Mg, for example).

\subsection{Distributions}

\begin{figure}
\centerline{
\psfig{file=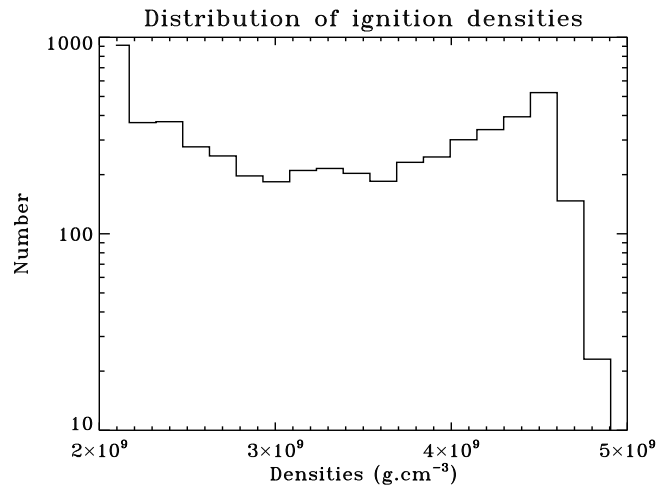,angle=90,width=9cm}
}
 \caption{Histogram of the central density at ignition for our 5576
SNe~Ia progenitors.}
\label{hrhoc}
\end{figure}

  We use the results of a binary population synthesis that produced
  5576 SNe~Ia progenitors out of 10$^7$ initial systems with
  $\alpha_{\rm CE}=\alpha_{\rm th}=1$ \citep[see][ section
  4]{HP04}. Note that the WD+red giant branch channel is not included
  in their study: this may affect the distributions.  We take the age
  when the WD reaches the Chandrasekhar mass and the initial mass of
  the WD as parameters $(t_{\rm a},M^{\rm i}_{\rm WD})$ which we
  interpolate linearly on our grid of models to get the central
  ignition density $\rho_{\rm i}$ for each of these systems. We
  display the histogram of $\rho_{\rm i}$ in figure~\ref{hrhoc}.

\begin{figure}
\centerline{
\psfig{file=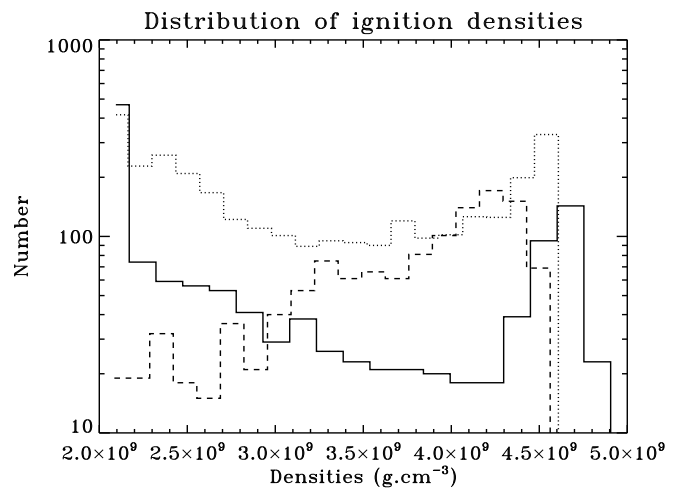,angle=90,width=9cm}
} \caption{Histogram of the central density at ignition for ages
greater than 0.8~Gyr (solid), ages between 0.4
and 0.8~Gyr (dotted) and ages lower than 0.4~Gyr
(dashed).  }
\label{hagerhoc}
\end{figure}

  The very sharp peak at the lowest density reflects the fact that
  accretion proceeds on longer than thermal diffusion time scales for
  the lowest initial masses $M^{\rm i}_{\rm WD}$. The secondary peak
  at high density reflects the shape of the ignition curve due to
  electron screening at high density. It is tempting to associate each
  one of these peaks to the clustering of either Branch-normal SNe~Ia
  or 91bg like events. But computations of the outcome of the
  explosions for each of our models need to be performed before we can
  assess the relation between ignition densities and
  observational properties of the explosion.  The shape of the
  distribution of $M^{\rm i}_{\rm WD}$ is rather flat and does not
  influence the overall probability distribution of $\rho_{\rm i}$.
  However, as noticed by \cite{HP04}, $M^{\rm i}_{\rm WD}$ is slightly
  correlated with $t_{\rm a}$ with lower ages giving rise to higher WD
  masses on average.  As a result more relative weight is given to the
  high density peak for younger systems (see figure~\ref{hagerhoc}).

\section{Composition effects}

  In this section we discuss the effects of the composition of the
  white dwarfs.  The initial mass and initial metallicity of their
  progenitors can vary independently but there can be a correlation
  between initial mass of the progenitor and that of the white dwarf
  at the end of the cooling phase.

\subsection{Initial C/O ratio}

\begin{figure}
\centerline{
\psfig{file=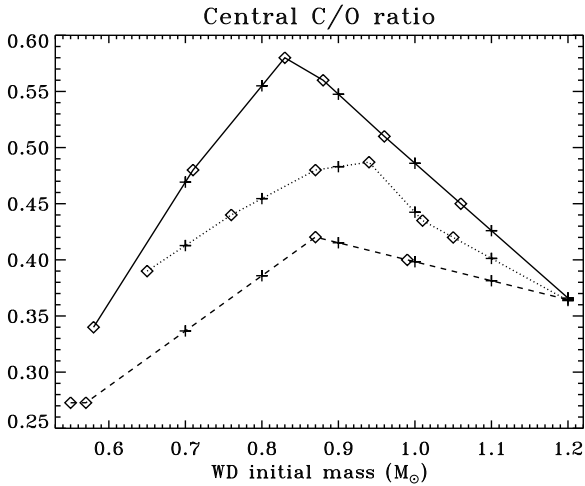,angle=90,width=9cm}
} \caption{C/O ratio for \citet{U99}, $Z=0.02$ (solid) and Z=0.001
(dotted) and for \citet{D01}, Z=0.02 (dashed). Diamonds are the values
computed by these authors, '+' signs are our interpolated values to 
compute the initial models.}
\label{co}
\end{figure}

  In the previous section all our results are based on central C/O
  ratios taken from \citet{U99}. However \citet{D01} have found
  significantly different values for the same solar metallicity as
  shown in figure~\ref{co}. This is probably due to their use of
  different $^{12}$C$(\alpha,\gamma)^{16}$O reaction rates with
  \citet{U99} using 1.5 times the rate of \citet{CF88} whereas
  \citet{D01} use a higher rate given by \citet{CF85}. We compute here
  a grid of models using central C/O values from \citet{D01} for a
  fixed cooling age of $t_{\rm a}=0.4$~Gyr and the same
  range of initial masses as previously.
  
\begin{figure}
\centerline{
\psfig{file=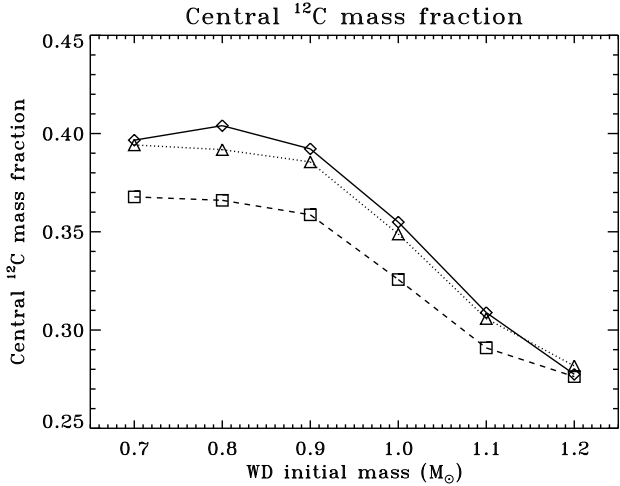,angle=90,width=9cm}
} \caption{Central carbon mass fraction at ignition for each initial
model of figure~\ref{co}. Diamonds (solid line) are our reference run
(see figure~\ref{xc}), triangles (dotted line) are for Z=0.001
\citep{U99} and squares (dashed line) are Z=0.02 \citep{D01}.}
\label{f3xc}
\end{figure}

  The resulting physical conditions at ignition are virtually
  identical to the runs described in the previous section. The only
  slight change is for the central carbon mass fraction which changes
  as shown in figure~\ref{f3xc}. This change only reflects the mixing
  in the convective core of the initial composition profile with the
  accreted matter.

\subsection{Metallicity effects}

  As shown in figures \ref{co} and \ref{f3xc} a change in metallicity
  is very much like a change in the initial C/O ratio, except that the
  magnitude of the effect is even less. In this respect, the
  metallicity effects can be strictly decoupled from the evolution in
  the C-flash and computed as in \citet{T03}. However, as noted by
  \citet{L05a}, electron captures could change the neutronisation in
  the course of the C-flash.  Whether or not the initial amount of
  $^{22}$Ne has an effect on the detailed nucleosynthesis during the
  C-flash can be addressed with a more extensive nuclear network
  than used in this study.

\section{Binary evolution effects}\label{binary}

  The results of previous sections strongly rely on the assumption of
  a permanent Hachisu wind, which allows us to neglect the binary
  evolution during the accretion process. Here we examine the effect
  of varying the accretion rate history.

\subsection{End of the wind phase}
\label{windend}

  From figure~1 of \citet{HP04} we find that the mass transfer rate
  decays exponentially after the wind has stopped. We hence model the
  mass transfer rate as
\begin{equation}
\dot{M}=\dot{M}_{\rm cr}(M_0)-\frac{M_{\rm WD}-M_0}{t_0}
\end{equation}
  where $M_0$ is the mass of the WD when the wind stops and $t_0$ is
  the time scale for the decay. The value of $t_0$ varies from $2.7\times 10^5\,$yr
  to $9\times 10^5\,$yr in different panels of figure~1 of
  \citet{HP04} and we take $6\times 10^5\,$yr as a characteristic
  value. We then use $\dot{M}_{\rm WD}=\eta_{\rm He} \dot{M}$.

  For three different masses $M^{\rm i}_{\rm WD}=0.8$, 1.0 and 1.2 M$_\odot$
  and $t_{\rm a}$=0.8~Gyr we compute the evolution after the end of the wind
  phase for $M_0=1.1$, 1.2 and 1.3~M$_\odot$ (except for $M^{\rm i}_{\rm
  WD}=1.2\,$M$_\odot$ for which $M_0=1.25$, 1.3 and 1.35~M$_\odot$).
  Only the central density and the convective core mass at ignition
  are found to change significantly compared to the models of section 2.

\begin{figure}
\centerline{
\psfig{file=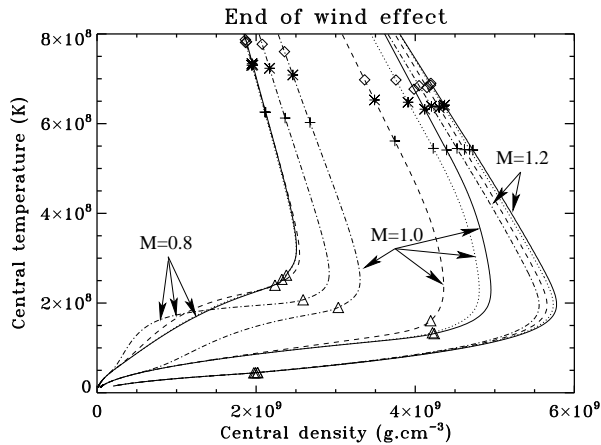,angle=-90,width=9cm}
} \caption{Central density-temperature evolution for three different
initial WD masses: 0.8 (left), 1.0 (middle) and 1.2~M$_\odot$
(right). Solid lines are the reference runs with a Hachisu wind always
on. Other lines stop the wind when the mass reaches three different
critical values of $M_0$ (see text). Labels are dotted, dashed and
dash-dotted for decreasing values of $M_0$.}
\label{macc}
\end{figure}

  Figure~\ref{macc} sums up the results for the $\rho_{\rm c}-T_{\rm
  c}$ tracks.  The slope of a given track in the $\rho_{\rm c}-T_{\rm
  c}$ diagram depends on the ratio of the rate of change of density
  and temperature. For a degenerate WD the density is directly linked
  to the total mass and hence the central density reacts immediately
  to any change in mass. On the other hand, the central temperature
  reacts to a change in accretion at the edge of the WD with a delay
  of the order of the thermal diffusion time scale $t_{\rm d}$ across
  the WD.  When the accretion rate slows down, the rate of change of
  the central density slows down immediately while the temperature
  continues to increase at the same rate: this steepens the slope of
  the $\rho_{\rm c}-T_{\rm c}$ track.  For the lowest masses (left in
  figure~\ref{macc}), the accretion takes place over a longer time
  than $t_{\rm d}$.  Hence global thermal balance is soon reached again and
  because the accretion rate is smaller this leads to a lower
  temperature for a given central density. The final result is a
  slightly higher ignition density.  For intermediate masses (middle in
  figure~\ref{macc}), global thermal balance was not realised at all
  because the accretion rate was too fast.  Hence the temperature was
  below the temperature the star would reach at global thermal
  balance. Slowing down the rate now gives time for the star to adjust
  and its temperature increases, which leads to lower central
  densities at ignition. For the highest masses (right in
  figure~\ref{macc}), there is not enough time for the star to react:
  the effects of slowing down the accretion rate do not reach the
  central thermal state of the star.

  A more significant difference exists for the mass of the convective
  core at the time of ignition. Indeed, the lower accretion rates
  yield lower temperatures at the edge of the core.  Because the outer
  boundary of the convective core is determined by the point where the
  inner adiabatic temperature profile meets the outer temperature
  profile, the convective core is bigger for lower accretion rates at
  a given central temperature. We find an increase of between 0.05 and
  0.2~M$_\odot$ in the convective core mass at ignition when the end of
  the wind phase is modelled. 

  We also computed the evolution of a WD using $M_0=1.12\,$M$_\odot$
  and $t_0=9.15\times 10^5\,$yr which gives a very accurate fit to the
  mass transfer rate for panel (a) of figure~1 of \citet{HP04}. But
  modelling the end of the wind phase actually makes only very little
  difference in this case due to the low initial mass of the WD
  (0.75~M$_\odot$). The only significant change was an increase of
  about 0.07 M$\odot$ in the convective core mass at ignition.

\subsection{Varying the critical accretion rate}

  To account for uncertainties in the critical accretion rate in the
  Hachisu wind model \citep[see the discussion in][]{HP05}, we ran the sequence of
  masses for $t_{\rm a}=$0.4~Gyr with twice the value provided by
  equation (\ref{mcr}). The effect turns out to be quite small, with
  less than a 10\% variation. WDs of lowest initial masses have a slightly
  lower $\rho_{\rm i}$ because the global thermal balance for a higher energy
  input rate yields higher temperatures. The effect is reversed at
  high initial masses because the shorter accretion times leave less
  time for the star to reach global thermal balance, so that the rise in
  central temperature is even more delayed and higher densities are
  reached at ignition.




\subsection{Qualitative results}

  To properly include the end of the wind would require modelling the
  stellar evolution of the secondary along with the WD.  We do not do
  this explicitly here.  However, a decrease of the accretion rate at
  the end of the wind phase as in section \ref{windend} mainly lowers
  the ignition density for WDs of intermediate initial masses. This
  suggests that binary evolution effects depopulate intermediate
  ignition densities to the benefit of the lowest densities.  Hence
  the probability distribution function of ignition densities should
  have a sharper peak at high density and a broader peak at low
  density. The broadening of the low-density peak would also be
  emphasised by the slight increase in $\rho_{\rm i}$ for lowest
  initial WD masses in systems that terminate their wind early on.



\section{Discussion}

  The main uncertainty in our model comes from the Hachisu wind model
  (discussed in section~\ref{binary}) but our results seem to hold
  even for significant variation in the accretion history.  In
  section~4 we showed that, although the initial C/O ratios are not
  well known, they do not affect our results very much.

  The main caveat of our study is that we do not include electron
  captures while our models go to quite high densities.  In the future
  we hope to be able to include the convective feedback as in
  \cite{L05a} but until now technical difficulties have prevented us
  \citep{L05b}.  If there was indeed a negative feedback on
  convective velocities, the mixing would be less efficient.
  Consequently the abundance changes would be enhanced and the
  explosion would probably take place sooner, at higher densities.  In
  particular electron captures would increase the neutronisation
  during the C-flash which may have dramatic consequences for the
  outcome of the explosion \citep{T03}.  Less homogeneous abundance
  profiles at the time of the explosion may also have consequences for
  the way the flame propagates because it goes through a medium
  increasingly concentrated in fresh fuel.  Finally electron captures
  would tend to slightly increase the density compared to our models
  but this would probably be a small effect.

  The high densities encountered in our models raise the question of
  whether the star will undergo an electron-capture supernova type of
  event leading to the formation of a neutron star rather than a
  thermonuclear explosion. However \cite{G05} indicate that a
  thermonuclear explosion is very likely given the high C mass
  fractions in our models.

  Accreting WDs are almost certainly strongly rotating. \citet{YL04}
  have shown how much influence this could have on the fate of the WD
  and how this could account for part of the type Ia
  diversity. However rotation is directly linked to the accretion so
  that it probably does not generate a new independent
  parameter. Our study has shown, at least in some cases, that the
  convective core is likely to reach the strongly differentially
  rotating region that helps support the star to higher than
  Chandrasekhar masses. Rotation might then help to increase the
  influence of binary evolution effects that seem secondary in this
  study.

  Finally this study is valid for the single degenerate channel
  only. If a Hachisu wind holds as well for the accretion of a WD on to
  another WD, we may apply part of the results presented here but in
  this case the C/O ratio of the accreted matter can be different from
  1. However the complete disruption of the accreted WD is likely to
  lead to a picture quite different from the Hachisu wind.


\section{Conclusions}

  We link the properties of single degenerate type Ia progenitors at
  the time of the ignition to results of a binary population synthesis
  thanks to successful models of the C-flash.

  We show that there is a large range of possible ignition densities
  whose distribution reflects the properties of the accretion on to
  the WD.  Furthermore, almost any other property of the WD at the
  time of the ignition is correlated to the ignition density: this
  makes it a one parameter family of models which considerably
  tightens the range of possible initial conditions for explosion
  models.  The convective time scales, amount of C-burning and central
  temperatures are functions of the central density at ignition. The
  central C mass fraction is better correlated to the initial mass of
  the WD progenitor.  The core mass at ignition can be modified
  according to how the mass transfer proceeds (this implies a
  sensitivity to the initial separation of the binary).  These last
  two parameters may hence be regarded as slightly independent of the
  ignition densities, although they do not have a large range of
  variation and are hence probably less essential than the density for
  the explosion.

  We have investigated several criteria that determine when the star
  would explode.  Although the actual dependence on density of the
  temperature, convective state and amount of C-burning are sensitive
  to the criterion chosen, we note that their correlations to density
  are still tight and the distribution of ignition densities does not
  change.

  We show that the metallicity has almost no effect on the C-flash
  phase.  Hence this parameter is independent of the central density
  at ignition.  Because both have been shown to have a big effect on
  the explosion outcome, we postulate that ignition density and
  metallicity are the two main parameters responsible for the
  diversity of type Ia supernovae. Metallicity has a linear effect on
  the peak brightness of type Ia supernovae with low metallicity
  yielding brighter $M_{\rm peak}$ \citep{T03,T05,M05}. However the effect
  of density on $M_{\rm peak}$ is still unclear: hydrodynamics of the
  explosion suggest that more mass is burnt to nuclear statistical
  equilibrium for higher densities but explosion nucleosynthesis
  \citep{T04} predicts that more stable elements are produced at
  higher densities because of electron captures.
  
  If density is the primary parameter and metallicity is the secondary
  parameter, then metallicity changes could shift the Phillips
  relation. On the other hand if metallicity is the primary parameter
  and density is the secondary parameter there should be less change
  in the Phillips relations, unless metallicity strongly affects the
  $M^{\rm i}_{\rm WD}$ distribution as postulated by \citet{U99b}.
  Finally, density {\it and} metallicity might be degenerate and both
  be primary parameters. The chain of computations from the ignition
  conditions to the light curves via explosion models and explosive
  nucleosynthesis (including electron captures) must be completed
  before we can theoretically address the effect of each parameter and
  uncover the physical nature of the Phillips relation. Observations
  of correlations with host galaxy metallicities \citep{Ga05} and
  comparison of type Ia luminosity functions with the metallicity or
  ignition density distributions will also give important clues.

\section*{Acknowledgements}
 This work was mainly funded through a European Research \& Training
Network on Type Ia Supernovae (HPRN-CT-20002-00303). It was also
supported in part by the Chinese National Science Foundation under
Grant Nos. 10521001 and 10433030 (Z.H.).


\begin{thebibliography}{}

 \bibitem[Caughlan et al.\ (1985)]{CF85} Caughlan G. R., Fowler W. A.,
 Harris, M. J., Zimmerman B. A. 1985, At. data Nucl. Data Tables 32,
 197

 \bibitem[Caughlan \& Fowler (1988)]{CF88} Caughlan G. R., Fowler
 W. A. 1988, At. data Nucl. Data Tables 40, 283

 \bibitem[Domínguez et al.\ (2001)]{D01} Domínguez I., Höflich P., Straniero
 O.  2001, ApJ 557, 279

 \bibitem[Dorfi \& Drury(1987)]{DD87} Dorfi E. A., Drury L. O'C. 1987,
 J. Comput. Phys. 69, 175

 \bibitem[Eggleton(1971)]{E71} Eggleton P. P. 1971, MNRAS 151, 351

 \bibitem[Eggleton et al.\ (1989)]{E89} Eggleton P. P., Tout C. A.,
 Fitchett M. J. 1989, ApJ 347, 998

 \bibitem[Eggleton et al.\ (1990)]{E90} Eggleton P. P., Fitchett
 M. J., Tout C. A. 1990, ApJ 354, 387

 \bibitem[Gallagher et al.\ (2005)]{Ga05} Gallagher J. S., Garnavich
 P. M., Berlind P., Challis P., Jha S., Kirshner R. P. 2005, ApJ 634, 210

 \bibitem[García-Senz \& Bravo(2005)]{GB05} García-Senz D. Bravo
 E. 2005, A\&A 430, 585 \bibitem[Hachisu et al.\ (1996)]{H96} Hachisu
 I., Kato M., Nomoto K. 1999, ApJ 522, 487

 \bibitem[Gutíerrez et al.\ (2005)]{G05} Gutíerrez J., Canal R.,
 García-Berro E. 2005, A\&A 435, 231

 \bibitem[Han \& Podsiadlowski(2004)]{HP04} Han Z., Podsiadlowski
 Ph. 2004, MNRAS 350, 1301

 \bibitem[Han \& Podsiadlowski(2005)]{HP05} Han Z., Podsiadlowski
 Ph. 2005, MNRAS {\it submitted}

 \bibitem[H\"oflich \& Stein(2002)]{HS02} H\"oflich P., Stein J. 2002,
 ApJ 568, 779

 \bibitem[Hurley et al.\ (2000)]{H00} Hurley J. R., Pols O. R., Tout
 C. A. 2000, MNRAS 315, 543
  
 \bibitem[Iapichino et al.\ (2005)]{I05} Iapichino L., Brüggen M.,
 Hillebrandt W., Niemeyer J. C., A\&A {\it accepted}

 \bibitem[Itoh et al.\ (1992)]{I92} Itoh N., Mitake S., Iyetomi H.,
 Ichimaru S. 1992, ApJ 395, 622

 \bibitem[Lesaffre et al.\ (2004a)]{L04} Lesaffre P., Chièze
 J.-P., Cabrit S., Pineau des Forêts G. 2004, A\&A 427, 147

 \bibitem[Lesaffre et al.\ (2004b)]{L05b} Lesaffre P., Tout C. A.,
 Stancliffe R. J., Podsiadlowski Ph. 2004, Memorie della Societa
 Astronomica Italiana 75, 660

 \bibitem[Lesaffre et al.\ (2005)]{L05a} Lesaffre P., Podsiadlowski Ph.,
Tout C. A. 2005, MNRAS 356, 131


 \bibitem[Mazzali \& Podsiadlowski(2005)]{M05} Mazzali P., Podsiadlowski
 Ph., ApJ {\it submitted}

 \bibitem[Phillips(1993)]{P93} Phillips M. M. 1993, ApJ 413, L105

 \bibitem[Pols et al.\ (1995)]{P95} Pols O. R., Tout C. A.,
 Eggleton P. P., Han Z. 1995, MNRAS  274, 964

 \bibitem[Röpke \& Hillebrandt(2004)]{RH04} Röpke F., Hillebrandt
 W. 2004, A\&A 420, L1

 \bibitem[Reinecke et al.\ (2002)]{R02} Reinecke M., Hillebrandt W.,
 Niemeyer J.C.  2002, A\&A 391, 1167

 \bibitem[Umeda et al.\ (1999a)]{U99} Umeda H., Nomoto K., Yamaoka H.,
 Wanajo S. 1999, ApJ 513, 861

 \bibitem[Umeda et al.\ (1999b)]{U99b} Umeda H., Nomoto K., Kobayashi
 C., Hachisu I., Kato M. 1999, ApJ 512, L43

 \bibitem[Timmes et al.\ (2003)]{T03}   Timmes  F. X.,   Brown  E. F.,
 Truran J. W. 2003, ApJ 590, 83

 \bibitem[Tonry et al.\ (2001)]{T01} Tonry J., The High-Z Supernova
 Search Team 2001, ASPC 245, 593
	
 \bibitem[Travaglio et al.\ (2004)]{T04} Travaglio C., Hillebrandt W.,  Reinecke M.,
Thielemann F.-K. 2004, A\&A 425, 1029

 \bibitem[Travaglio et al.\ (2005)]{T05} Travaglio C., Hillebrandt W.,
 Reinecke M. 2005, A\&A 443, 1007

 \bibitem[Woosley et al.\ (2004)]{W04} Woosley S., Wunsch S., Kuhlen M.
 2004, ApJ 607, 921

 \bibitem[Wunsch \& Woosley(2004)]{WW04} Wunsch S., Woosley S. 2004,
 ApJ 616, 1102

 \bibitem[Yoon \& Langer(2004)]{YL04} Yoon S.-C., Langer N. 2005, A\&A 419, 623

\end{thebibliography}
\end{document}